\documentclass[11pt]{amsart}
\pdfoutput=1

\usepackage[utf8]{inputenc}

\usepackage[margin=3.25cm]{geometry}

\usepackage{amssymb, amsmath, amsthm, amsfonts}
\usepackage{graphicx}
\usepackage{color}
\usepackage{mathrsfs}
\usepackage{physics}
\usepackage{mathtools}
\usepackage{tikz}
\usepackage{faktor}
\usepackage{tikz-cd}
\usepackage{bm}
\usepackage[hidelinks]{hyperref}
\usepackage{csquotes}

\DeclareRobustCommand{\SkipTocEntry}[5]{}

\newtheorem{theorem}{Theorem}

\newtheorem{corollary}{Corollary}
\newtheorem{defn}{Definition}
\newtheorem{prop}{Proposition}

\theoremstyle{remark}
\newtheorem{remark}{Remark}
\theoremstyle{remark}
\newtheorem{example}{Example}

\usepackage[UKenglish]{isodate}
\usepackage[UKenglish]{babel}

\newcommand{\R}{\mathbb{R}}

\DeclareMathOperator{\E}{\mathbb{E}}
\newcommand{\lag}{\mathcal{L}}

\DeclareMathOperator*{\argmin}{\text{arg\,min}}

\renewcommand{\grad}{\nabla}
\newcommand{\tra}{\text{tra}}
\newcommand{\traemph}{\emph{tra}}
\newcommand{\g}{\mathfrak{g}}

\newcommand{\Exp}{\text{Exp}^\times}
\newcommand{\Expemph}{\emph{Exp}^\times}
\DeclareMathOperator{\U}{\mathcal{U}}

\title{entropy-maximising diffusions satisfy a parallel transport law}

\author{Dalton A R Sakthivadivel}
\address{\parbox{\linewidth-12pt}{Department of Mathematics, Department of Physics and Astronomy, Stony Brook University, Stony Brook, NY, 11794-3651, USA}}
\address{VERSES Research Lab, Los Angeles, CA, 90016, USA}
\email{dalton.sakthivadivel@stonybrook.edu}
\urladdr{https://darsakthi.github.io}
\date{\today}

\let\oldtocsection=\tocsection

\let\oldtocsubsection=\tocsubsection

\renewcommand{\tocsection}[2]{\hspace{0em}\oldtocsection{#1}{#2}}
\renewcommand{\tocsubsection}[2]{\hspace{19.5pt}\oldtocsubsection{#1}{#2}}

\begin{document}

\maketitle

\begin{abstract}

We show that the principle of maximum entropy, a variational method appearing in statistical inference, statistical physics, and the analysis of stochastic dynamical systems, admits a geometric description from gauge theory. Using the connection on a principal $G$-bundle, the gradient flow maximising entropy is written in terms of constraint functions which interact with the dynamics of the probabilistic degrees of freedom of a diffusion process. This allows us to describe the point of maximum entropy as parallel transport over the state space. In particular, it is proven that the solubility of the stationary Fokker--Planck equation corresponds to the existence of parallel transport in a particular associated vector bundle, extending classic results due to Jordan--Kinderlehrer--Otto and Markowich--Villani. A reinterpretation of splitting results in stochastic dynamical systems is also suggested. Beyond stochastic analysis, we are able to indicate a collection of geometric structures surrounding energy-based inference in statistics.

\end{abstract}

\date{\today}


\tableofcontents

\section{Introduction}

The configurations of statistical fields almost universally tend to maximise their entropy in classical statistical field theory. Examples of what is called the principle of maximum entropy abound in physics \cite{jaynes} and functional analysis \cite{JKO, villani}. Its utility is undisputed, yielding approaches at the edge of our understanding of statistics such as machine learning. Given this, however, we still have very little rigorous understanding of the effectiveness of machine intelligence and other complex statistical algorithms \cite{terrence, lin}, which seem to compute the solutions to arbitrarily complicated dynamical systems as data-generating processes, without ever taking a single integral. Indeed, the partial differential equations determining the evolution of systems with probabilistic degrees of freedom are often non-linear, non-stationary, or otherwise difficult to handle analytically. 

Such algorithms instead take inspiration from variational calculus, approaching these problems by inducing a gradient flow on some functional of the state of the system. For many problems, this procedure is either constructed tautologically\textemdash the functional extremised is some metric on the solution space, yielding the distance between a given solution and a solution known to have desired properties\textemdash or, maps onto an underlying principle, like the minimisation of the kinetic action \cite{lecun}, the maximisation of entropy \cite{jaynes}, or the extremisation of other energy functionals \cite{afepfapp}. In this context, the principle of maximum entropy is particularly interesting, because several other statistical methods are cases of maximum entropy under individual constraints \cite{ebl}, and maximising entropy is able to produce exact solutions to an enormous range of processes \cite{presse}.

Whilst these variational methods are powerful, this is apparently where the analogy between inference and functional optimisation ends: much of statistical inference is, at the moment, not properly functional analysis, and relies on simpler techniques from multivariable differential calculus like `backpropagation.' Some other insight must be needed to make sense of the success of these frameworks. Taking the dynamical view of inference, wherein we can consider the act of characterising a process as solving some kind of differential equation, builds a bridge from the purely analytic view to a complementary geometric view. The idea of state spaces with non-trivial potential functions\textemdash and dynamics in such spaces\textemdash leads us to gauge theory. The success of gauge theory as a \emph{physical} device is unparalleled; physically, gauge symmetries often unify several different phenomena in a theory into a single, coherent, descriptive law or equation \cite{witten}, and can be used to describe much of the known universe. Mathematically, the geometry behind gauge theory has become an ideal mathematical language in which to describe the dynamics of physical fields. Gauge theories use generalisations of dynamical systems rooted in geometry, like the covariant derivative or the equivalence principle, to relate dynamics in a space to the structure of the space \cite{bleek, nakahara}. The variational principles involved in this geometric approach extend the gradient flows already in use throughout statistical inference and PDE analysis, as in the study of geodesics of pseudo-Riemannian manifolds in general relativity. Moreover, relying on sections and the geometry of tangent vector fields removes troublesome analytic elements, like the exact equations of motion generating the field. The incorporation of physical information about the inference process introduces some extra structure to the state space, in terms of a prior specification of what states are acceptable. Taking highly constrained states as unlikely to be sampled, this naturally shapes the variational dynamics of inference as it solves a PDE for the probability density over states. This approach maps on to gauge theory well, in that it also avoids analytic difficulty by geometric means.

Taking inspiration from the mathematics and physics behind these questions about the effectiveness of probabilistic inference in both statistics and PDE theory, we provide some geometric preliminaries, and then apply the indicated gauge-theoretic insights to determine how maximum entropy\textemdash and inference more generally\textemdash operate in the state space of a diffusion process. In Theorem \ref{J-is-gauge-field-thm} we identify a gauge symmetry in maximum entropy, explaining a more general observation (put forth in Theorem \ref{par-trans-at-max-ent-thm}) that the stationary point of the entropy functional appears to describe a surface on which parallel transport is performed. This naturally occurs on horizontal level sets of the surface, and indeed, these curves are horizontal lifts of the integral curves of the kernel of a local gauge field (given as a pullback connection one-form). Theorem \ref{main-result} is the main result of this paper, which states that finding the probability density associated to some stochastic dynamical system is equivalent to solving a much simpler, linear problem describing the geometry of the surface we identify as the probability density, as it is determined by a constraining potential function. Some elementary knowledge of both gauge theory and statistical inference is assumed; the latter is because we are motivated by asking how statistical algorithms solve diffusion processes in a potential. 



\section{Preliminaries}

We speak\textemdash somewhat informally\textemdash of an object field, or an object-valued field theory, as a continuous (at minimum) function of an $n$-dimensional input space, assigning values of some object to a base space. For instance, a quantum field theory is a field of quantum states for a particular sort of particle over space-time. A field theory typically consists of the object field, as well as other important data about the field's behaviour, like the action. A field configuration is a particular set of states instantiating the abstract field. Later, we will refer to this as a choice of section. The configuration of a physical field arises from an action functional, whose extremal points lead to the equations of motion for the state of the field. This functional is typically a full physical characterisation of a field. 

Physically, a gauge theory is a field theory with a free choice of `gauge,' a quantity which comprises an arbitrary degree of freedom in the theory \cite{rubakov}.  If the action, or the variation of the action, is invariant under the value of any such gauge, then many distinct field configurations may be action-wise equivalent; hence, functionally and physically, there is no difference between these configurations. Whilst the action is invariant, the field itself is typically \emph{covariant}, meaning it couples to the gauge function such that they change together. Such a redundancy in the theory\textemdash that is, physically equivalent fields which have redundant individual descriptions, all of whom are related by gauge transformations\textemdash arises from a local symmetry of the action.

Let $X$ be a smooth manifold of dimension $n$. We also equip $X$ with an atlas $\U = \{(U_\alpha, \psi_\alpha): \alpha \in I\}$ of local patches. For brevity, we denote a coordinate in any such patch by $x$, which is generically a point $\psi_\alpha(\xi) \in \R^n$; it refers to a vector of pulled back coordinate functions $[x^1(\psi_\alpha\circ\xi), \ldots, x^n(\psi_\alpha\circ\xi)]$ at some point $\xi \in X$ contained in the patch. Gauge theory occurs in a principal $G$-bundle over $X$, denoted $P\xrightarrow{\pi_P}X$. This is a fibre bundle with base $X$ wherein the typical fibre is isomorphic to a Lie group $G$, and there is a right action of $G$ on the total space $P$. Let $F$ be a vector space over a field $K$. There is, in general, a vector bundle $E$ associated to $P$, such that each fibre $E_x$ is a vector space with a symmetry $G \cross F \to F$. The bundle $P \cross_G F$, which is what we denote by $E$, can be constructed by the quotient 
\[
\faktor{P\cross F}{G}
\]
which takes coordinates $(p, f)$ and assigns each to an equivalence class such that $(pg, f) = (p, g^{-1} f)$. In other words, these are orbits of $G$.

An $F$-valued field is an $F$-valued function, $f : U_\alpha \to F$, extended over the input space $X$. Taking a section $s : U_\alpha \to E$, $x \mapsto (x, f(x))$, we have a function whose image is valued in $F$; hence, a field is given by a local section of $E$. By construction, each such pair is a point in $E$ associated to a set of group elements, representing the equivalence class $[s(x), \sigma(x)]$ for $\sigma : U_\alpha \to P$. Such a vector bundle has structure group $G$ but typical fibre $F$, and thus records a space which is invariant under the action of a faithful, linear representation of $G$\textemdash that is, the transition functions re-framing a local section are $G$-valued. Taking the isomorphism on fibres induced by $\sigma : U_\alpha \to P$ leads us to the previous function $s : U_\alpha \to F$, coupling the field to the gauge potential. Suppose the group $G$\textemdash identically, $P_x$\textemdash is a symmetry group of some action functional. Then, as a $g$-valued function on points $x \in X$, a choice of gauge is a choice of (usually local) section.

On any manifold $M$, including the total space of a fibre bundle, there exists a bundle of fibres consisting of all the tangent vectors to a point on $M$. This vector bundle, $TM$, has fibres $T_m M$ with elements $v(m)$, such that a section $M \to TM$ is a vector field $(m, v(m))$. The same is true of $P$, in that there is a bundle $TP$ with tangent vectors $v(p) \in T_p P$. A choice of a codimension one horizontal tangent space, such that $T_pP$ consists of two complementary subspaces $V_pP$ and $H_pP$, is known as a \emph{connection}. The connection itself is a one-form $\omega_p$ projecting tangent vectors into $V_pP$, such that $H_p P$ is given by $\ker(\omega_p)$. The vertical tangent space $V_p P$ can be intuited as coinciding with the fibre sticking up from the base; indeed, the fact that the fundamental vector field at $p$ generated by elements of the Lie algebra is contained in $G = P_x$ is important for the definition of the connection. The horizontal space $H_p P$, on the other hand, can be thought of as a space from which we generate vector fields on the base manifold, given by the kernel of $\omega$. In particular, the splitting of the tangent space into vertical and horizontal components is such that any individual vector $v(p) \in T_p P$ decomposes into 
\[
v(p) = \omega_p(v(p)) + v^H(p)
\]
where $v^H(p)$ is a horizontal lift of a vector $v(x) \in T_{\pi_P(p)} X$\textemdash that is, a lift of some vector on the base below the point in question. This isomorphism arises from the splitting of the Atiyah sequence
\[
0 \to VP \to TP \to \pi_P^* TX \to 0 
\]
by $TP \simeq VP \oplus HP$, leading to a connection on the left and a lift on the right.\footnote{By exactness, $VP$ embeds into $TP$ and is then sent to zero to get the horizontal subbundle; indeed, $VP$ is the kernel of the horizontal projection map in $TP$. The pullback bundle $\pi_P^* TX$ associates to points in $P$ tangent vectors on $X$, such that the fibre and the vector share the same base point $x$ and the lifted vector is parallel to the vector on the base\textemdash that is, they both point in a horizontal `direction.' Specifying an isomorphism between the pullback bundle and the horizontal tangent space is the element of choice present.}
Likewise, the identification of the vertical component of the vector with the fibre means all that is left in the vector is parallel to the underlying vector, and placed up along the fibre by a choice of vertical component.

This suggest the first item of interest: the existence of a gauge symmetry implies a particular geometry on the tangent space, thus affecting the possible dynamics on $X$. As was noted, a free choice of gauge becomes a choice of local section $\sigma : U_\alpha \to P$, and the local gauge field on $X$, in fact a covector field, induced by the gauge potential is the pullback $\sigma^*\omega$. The gauge field itself has the interpretation of an internal potential field that a field interacts with, guiding its dynamics under any such choice of gauge (see, for instance, the exposition in \cite{baez} or \cite{baez2}). The connection on the principal bundle leads to an idea of parallel transport, which is the movement of a point along a curve such that the tangent vector to the curve is horizontal. That is, the tangent vectors along the curve do not themselves curve, transporting themselves parallel to tangent vectors along curves in the base. Moreover, parallel transport follows the connection by virtue of staying in $H_p P$. In this context, a connection can be regarded as a potential function, in that it gives us a choice of what counts as a horizontal tangent vector\textemdash and thus, a specification of what sort of velocity field a path in $P$ (or in $E$ via the covariant derivative) obeys. Thus, the parallel transportation of a curve along a section is really parallel on the connection. As such, parallel transport solves for the dynamics of a particle obeying a gauge potential. 

The induction of a covariant derivative of sections on $E$\textemdash a section that follows the geometry of the principal bundle, such that the integral curves of the resulting vector field are geodesics with respect to some potential generating the geometry\textemdash is equivalent to a
linear dynamical system. Observe the following: for a parallel section, 
\[
(\dd{} + \omega)s(T)  = 0.
\]
In other words, the derivative of a section follows the negative connection form exactly. From the perspective of the covariant derivative, the connection form is a correction term that keeps track of the varying choice of gauge across $x$ points. Since the connection is also roughly like the gradient of the potential, this can be seen as a generalisation of a dynamical system, $\dot{s} = -\grad F$. In fact, if the Lie algebra in which $\omega$ is valued consists of matrices, we have the system of ODEs
\begin{equation}\label{ode-par-trans}
\dd{s}^i_j(T) = -\omega^i_j(T) s
\end{equation}
for any directional vector $T$. Parallel transport is always the solution to this equation, no matter how complicated the potential identified with $\omega$ is, so long as $\omega$ has zero curvature\textemdash that is, its kernel contains all horizontal vectors, such that $E$ admits a covariantly constant section. It says how the integral curves of the covariant derivative should look, and under the above identification, gives the dynamics of a system following the potential it is subject to, as it is reflected in a deformation of the geometry guiding that solution. This is the key to the results presented here.

Using the objects defined here, we will build inference as a geometric theory of probabilistic dynamics.

\section{Stochastic dynamical systems maximise their entropy}

When we talk about a process, we generally refer abstractly to a function which produces a set of objects indexed by some input argument. A field theory is a process, by definition of a section. Let $X$ be the state space of some process. Also, let $\varphi : X \to X$ be a one-parameter subgroup of diffeomorphisms of $X$, i.e, a path consisting of $x$-coordinates indexed by some variable $t$ corresponding to the solution of a dynamical system. We will generally refer to the process by looking purely at its image, which is one such curve\textemdash that is, we forget the abstract map generating the trajectory, and just look at the sequence of states. This is akin to identifying a section $s$ with its image $f(x)$. Under this approach, a useful intuitive picture of a dynamical system is like a marble rolling about in its state space. We can forget about the system itself, and instantiate it in $X$\textemdash in other words, the existence of so-called `generalised coordinates' induces a sort of isomorphism between dynamics \emph{valued} in a space of states, and dynamics \emph{situated} in a space of states. This is also true of parallel transport as the generator of a horizontal lift.

\begin{defn}
A dynamical system on $X$ is a pair $(\varphi_0, v)$, or, a tangent vector field $v : X \to TX$ whose integral curve satisfies an initial condition $\varphi_0$. The solution to the dynamical system, $\varphi$, is this integral curve.
\end{defn}

The geometric view of a dynamical system is essentially that of a flow through some tangent vector field; a dynamical system is thus a differential equation of some character (possibly noisy or non-linear) whose solution is a list of $x$ values (states, or outputs) parameterised by some variable $t$ (an input). When said informally, we will sometimes identify `dynamical system' with the solution $\varphi$. When our system possesses one-dimensional orbits, or trajectories, it can be described as a path in the state space. The solution to the system is thus a subgroup of the group of diffeomorphisms $X$ under composition, parameterised by $t$, which was our function $\varphi$. This curve\textemdash the solution to the dynamical system\textemdash describes a process which generates the dynamics observed.\footnote{Recall we have made a distinction between a process, and a curve in $X$ serving as a description of that process by solving some equation of motion.} Taking a random process, where any particular state $\varphi_t$ comprising a solution is generated under some noisy dynamics, statistical inference can be characterised as a process which attempts to determine the probability measure over the state space. Thus, in estimating the probability of observing any particular state, we naturally solve the associated deterministic dynamical system for $\partial_t p$. This is what constitutes learning or inferring the process generating some observed states.

\begin{defn}\label{max-ent-def}
Let $\gamma(x, t)$, denoted $\gamma_t$ for brevity, be an $X$-valued random variable satisfying some stochastic differential equation, with labelled instances $\gamma_t = x_t$. Moreover, let the process generating $\gamma_t$ be stationary, such that $\gamma_t$ is distributed like $p(x)$ for every $t$, and let $J : X \to \R$ be a measurable scalar function. The Shannon entropy is the action functional
\begin{equation}\label{max-ent-action}
S[p; J] = -\int_X \ln\{p(x)\} p(x) \dd{x} - \sum_k \lambda_k \left(\int_X J_k(x) p(x) \dd{x} - C_k \right), 
\end{equation}
where any $C_k = \E[J_k(\gamma_t)]$ such that the final term is zero. 
\end{defn}
Without loss of generality, we typically write $\lambda J$ for a linear combination of constraint functions, $\sum_k \lambda_k J_k$. We have denoted this integral and the function $J$ as being over $X$, but will no longer do this to emphasise that both are local objects existing on an arbitrary patch on $X$.

The stationary point of this functional given a particular choice of $J$ can be found by the variation of \eqref{max-ent-action}, which is simply the Euler--Lagrange equation
\[
\pdv{p(x)} \big( - \ln\{p(x)\} p(x) - \lambda J(x) p(x) \big) = 0.
\]
This yields the solution $\exp\{ -\lambda J(x) \}$ in general. Various well-known results in the analysis of PDEs elaborate on how the gradient flow maximising constrained entropy is the Fokker--Planck equation\textemdash that is, $\exp\{ -\lambda J(x) \}$ is also the stationary solution to diffusion in a potential $J(x)$, whose dynamics describe the maximisation of entropy \cite{JKO, villani}. Here, $\lambda$ is a Lagrange multiplier, which we consider an arbitrary degree of freedom in the remainder of this paper. In that $\lambda$ is a force field enforcing the constraint, it can be handled as an effective field to be integrated out or neglected; it is certainly the case that solving for $\lambda$ merely gives constants, such as the partition function, or other parameters controlling the shape of the distribution at a finer level than just the constraint function.

The statistician's view of maximum entropy is generally that a constraint plays the role of a specification of the moments of a probability density $p$, in that we directly ask that the expectation of some quantity $J$ with respect to $p$ is some fixed quantity $C$ \cite{presse}. If $J(x) = x$ or $J(x) = x^2$, for instance, then the interpretation is even more obvious: placing a constraint is demanding that, whatever density is produced, it must have a mean (variance, respectively) of $C$. In a sense, this shapes the distribution by changing the statistics of possible samples \emph{a priori}. Taking the view of an optimisation theorist, meanwhile, the constraint is something like a penalty term that changes the point at which the maximum of $S$ occurs. Convex duality dictates that maximising constrained entropy is equivalent (under constraint qualification conditions) to maximising entropy on a constrained set of allowable $p(x)$'s \cite{vector-opt}. The introduction of this penalty is exhibited by changing $-\ln\{p(x)\} = 0$ to $-\ln\{p(x)\} = \lambda J(x)$, in turn changing the possible landscape of solutions to the minimisation of $S$\textemdash and indeed, directly changing the solution to the Euler--Lagrange equations. Let $H[p]$ be the entropy functional. For a vector field on the space of solutions $p(x;J)$ to satisfy the variation of \eqref{max-ent-action}, the existence of a constraint level set $\E[J] = C$ demands that at some point $p(x; J)$,
\[
\grad H[p] = \lambda \grad \E[J],
\]
which returns 
\begin{equation}\label{max-ent-lagrangian-eq}
-\ln\{p(x)\} = \lambda J(x)
\end{equation}
for the gradient in $L_2(\dd{P})$. This reflects that $\lambda$ is merely a velocity function on the curve, or, a non-singular transformation scaling the tangent vectors in the aforementioned vector field.

Interestingly, $J(x)$ itself can be read as a penalty on states $x$: the idea of setting a constraint function is that the mean of that function is some desired quantity, which admits a \emph{post hoc} interpretation of penalising states $x$ for whom $J(x)$ is too far from that mean. This can be simplified to say that the probability of a state is weighted in direct proportion to $-J(x)$, up to composition with some other function that ensures strict positivity whilst still respecting the inversion induced by the problem. The canonical choice of such a function is the exponential function, due to its well-known qualities as a group homomorphism\textemdash in particular, that it takes additive inverses to multiplicative inverses. Thus, we have justified the solution $\exp\{-J(x)\}$ backwards. The point of this paper is, in some sense, to see how instructive this geometric view of constraint functions can be.

\section{On the parallel transport of probability}\label{gauge-symmetry-section}





The key motivation for the approach in this paper is twofold: that there ought to be a duality between deformations of the geometry of a state space and constraints on the dynamics in that state space, in such a way that (i) maximum entropy translates between these two viewpoints, and moreover, that (ii) maximum entropy describes a density consisting of `particles' moving in the presence of a gauge field. Thus, much like parallel transport gives fields as horizontal lifts of paths on space-time into the total space of the bundle contingent on a chosen gauge field, maximising entropy to build a probability density should be described equivalently by transporting probability over states in parallel fashion. 

The latter point is exemplified in the following more general observation: the point in the space of probability densities where entropy is at its maximum is the density corresponding to parallel transport on contour lines of the function $J$.

\pagebreak

\begin{theorem}\label{par-trans-at-max-ent-thm}
Parallel transport occurs at the point $p(x)$ of maximum entropy, which directly solves the stationary Fokker--Planck equation via an integral relationship encoded in \eqref{max-ent-lagrangian-eq}.
\end{theorem}
\begin{proof}
Suppose we choose a constraint function $J$ and define a connection one-form $\dd{J} = \partial_i J(x) \dd{x}^i$, whose kernel is a distribution with integral curves $J(x^1, \ldots, x^n) = c$. The Euler--Lagrange equation satisfying $\argmin_{p}[S]$ is \eqref{max-ent-lagrangian-eq},
\[
-\ln\{p(x)\} - \lambda J(x) = 0,
\]
which is equivalent to the integral of \eqref{ode-par-trans}, the ODE for parallel transport along any such level set of $J$:
\begin{align*}
    \dv{x} \ln\{p(x)\} &= - \dv{x} \lambda J(x) \\
    \dd{ p(x)} &= -\lambda \,\partial_i J(x) \dd{x}^i p(x).
\end{align*}
Now, in the standard basis of $\R^n$, we get the classic ODE 
\begin{equation}\label{gradient-eq}
\partial_i p(x) = -\lambda \,\partial_i J(x) p(x).
\end{equation}
By the gradient theorem, this equation integrates a stationary Fokker--Planck equation with potential $J(x)$ and vector field $-\partial_i J(x)$; its own integral, and thus the solution to the Fokker--Planck equation, is \eqref{max-ent-lagrangian-eq}.
\end{proof}

This theorem gives a straightforward analytic reason for the equivalence of the solutions to parallel transport and maximum entropy, as well as the exponential form of those solutions. Indeed, the gradient flow maximising $S$ famously gives the stationary solution to the Fokker--Planck equation\textemdash however, the connection to an ODE whose integral curves generate level sets of the probability density has not previously appeared in the literature. Conversely, this result privileges the Shannon entropy as a `canonical' choice of entropy functional, in that at the stationary point of the Shannon entropy, parallel transport occurs; this is clearly not true of functionals like the Tsallis entropy. 

\begin{remark}\label{differences-rem}
Despite the similarity in appearance between \eqref{max-ent-lagrangian-eq} and \eqref{gradient-eq} being suggestive, this is evidently a coincidence: building a density as a surface consisting of integral curves of an involutive distribution is a purely local construction, exists in a state space and not a space of probabilities, and can be constructed for any $p$, whether or not $S[p;J]$ is extremal. The underlying principle\textemdash that constraining the shape of the surface $(x, p(x))$ defining our density is implemented in the same way as placing constraints on the point $p(x)$ at which entropy is maximised\textemdash explains this coincidence. That is, \eqref{gradient-eq} is the gradient of \eqref{max-ent-lagrangian-eq}, and \eqref{max-ent-lagrangian-eq} is the (functional) gradient of \eqref{max-ent-action}. Gauge symmetries and parallel transport have both appeared in previous work on statistical inference \cite{amari, friston} and statistical physics \cite{polettini}. These results, which are of independent interest, focus all the same on the dynamical aspects of flows, as opposed to the idea of covariance with constraints at the point of maximum entropy. As we stated, this is a key difference between the constructions.
\end{remark}

Theorem \ref{par-trans-at-max-ent-thm} is, however, na\"ive as written. The claim is only true if $J$ indeed has the interpretation of a connection in some bundle to which $p$ lifts, such that \eqref{max-ent-lagrangian-eq} truly corresponds to the solution of the parallel transport ODE. This is precisely where having a geometric interpretation of maximum entropy would be useful. 

An ideal geometric framework is gauge theory, for its stated dynamical aspects. Given we are agnostic to the solution of the field equation arising from the variation of \eqref{max-ent-action}, and given maximising constrained entropy is the same as maximising unconstrained entropy on a constraint manifold\textemdash here, a level set $\E[J(\gamma_t)] = C$ through the space of probability densities\textemdash there is a suggestion of the constraints being redundant. Indeed, referring to \eqref{max-ent-action}, the action is written explicitly in such a way that the constraints zero out and $S[p] = S[p ; J]$. Despite this apparent redundancy, the choice of gauge does change the solution to the maximisation of $S$; thus, the constraints resemble a choice of gauge coupling to a covariant `matter field,' $p(x)$. In this framework, we can also define a transformation rule for $p(x)$, deriving from a change in constraints. This redistributes the probability mass away from constrained regions, by acting on $p(x)$ with an exponential function. The result is consistent with re-maximising entropy under $J'$, or identically, changing the penalty on subsets of the support of $p(x)$. 

In the following theorem, we show invariance under an initial choice of $J$ and any such change in gauge explicitly.

\begin{theorem}\label{J-is-gauge-field-thm}
The maximum entropy action couples the Shannon entropy functional to the expected constraints such that the action of the scalar field $p(x)$ is gauge invariant.
\end{theorem}
\begin{proof}
Consider \eqref{max-ent-action}. Choose an arbitrary constraint function $J$, and set $C=0$ without loss of generality. We will show there is a gauge symmetry in maximum entropy, expressed by the invariance of $S$ under the transformation $p(x) \mapsto \exp{-\lambda J'(x)}p(x)$, and the resultant change to $\lambda J'(x) + \lambda J(x)$. These give
\[
S[p';J'] = - \int \ln\{e^{-\lambda J'(x)}p(x)\}e^{-\lambda J'(x)}p(x) \dd{x} - \int \big(\lambda J'(x) + \lambda J(x)\big) e^{-\lambda J'(x)} p(x) \dd{x}
\]
for $S$. A brief calculation shows the Lagrangian 
\[
\lag = - \ln\{e^{-\lambda J'(x)}p(x)\}e^{-\lambda J'(x)}p(x) - \big(\lambda J'(x) + \lambda J(x)\big) e^{-\lambda J'(x)} p(x)
\]
simplifies to
\begin{align}
\lag &= \big(\lambda J'(x)-\ln\{p(x)\}\big)e^{-\lambda J'(x)}p(x) + \big(-\lambda J'(x) - \lambda J(x)\big) e^{-\lambda J'(x)} p(x) \nonumber \\
&= \big(-\ln\{p(x)\} - \lambda J(x) \big) e^{-\lambda J'(x)} p(x) , \label{temp-eq}
\end{align}
the variation of which yields 
\[
\big(-\ln\{p(x)\} -  \lambda J(x) \big)e^{-\lambda J'(x)} = 0.
\]
Since $\exp{-\lambda J'(x)} > 0$ for all real-valued $\lambda J'(x)$, the above equation holds only when
\[
-\ln\{p(x)\} - \lambda J(x) = 0,
\]
and as such, $S[p';J']$ has the same root as $S[p;J]$. In new coordinates
\[
p'(x) = \exp{-\lambda J'(x)}p(x),
\]
the variation of \eqref{temp-eq} is
\[
-\ln\{p(x)\} - \lambda J(x) = 0,
\]
which is again identical to \eqref{max-ent-lagrangian-eq}. Hence $S[p';J']$ has the same root as $S[p;J]$ in both coordinate systems.
\end{proof}

Treating data with maximum entropy demands a specific set of constraints, but much like a gauge theory, it is not initially apparent that any such constraint matters\textemdash the initial choice of constraints was completely arbitrary, and could have just as well been an unconstrained problem with $J$ being the constant function equal to zero. Likewise, the transformation is a change in the choice of constraints to something just as arbitrary. Again, like a gauge theory, we generally fix a constraint (especially `at runtime,' that is, in \eqref{max-ent-lagrangian-eq} by choosing a particular $J$ function) and calculate a gauge covariant field equation\textemdash but, we want to have the flexibility to treat other constraints as equally valid in general, especially re-parameterisations of a fixed constraint. This is a generalisation of previous coordinate invariance arguments due to Jaynes, an essential part of Shore and Johnson's later axiom scheme for maximum entropy inference (see \cite{presse} for a review). Likewise, the redundancy of constraints enables us to do away with them; as such, we can use the entropy functional both before and after maximisation \cite{dill2, dill-reply} and still get meaningful results about a system (in the latter case, this yields thermodynamical quantities).

Since the free choice of gauge is formally a choice of section to a principal $G$-bundle with a $G$-equivariant associated vector bundle, we show the existence of both to introduce the geometric material we will leverage. 

\begin{defn}\label{exp-defn}
Let $\R_{> 0}$ denote the set of strictly positive real numbers. We define $\Expemph$ as a closed submanifold of $\R$ consisting of elements $\exp{q}$ for some $q\in \R$, whose image is in bijection with $\R_{> 0}$. It is easily verified that $\Expemph$ is closed under multiplication, has a multiplicative identity and inverses, and, that this multiplication commutes. The homomorphism $\exp{q'+q} = \exp{q'}\exp{q}$ thus defines $\Expemph$ as an abelian Lie subgroup of $(\R, +)$. Furthermore, this group is in fact a subgroup of $GL(1, \R)$, the multiplicative group of non-zero real numbers. Taking exponentiation in $\R$, then by general properties of the exponential map\textemdash namely, that $\exp_G\{\emph{Lie}(G)\} = G$\textemdash the Lie algebra of $\Expemph$ is isomorphic to $\R$.
\end{defn}

\begin{prop}\label{fibre-bundle-construction-prop}
There exists a real line bundle $E$ with section $s : x \mapsto (x, p(x))$ associated to the principal $\Expemph$-bundle $P$ with section $\sigma : x \mapsto \left(x, e^{J(x)}\right)$ such that $E$ is a representation of $\Expemph$.
\end{prop}
\begin{proof}
We construct both objects explicitly. Take an open cover of $X$ and an intersection $U_{ij} = U_i \cap U_j$ containing a point $x$. Suppose that, on any $x \in X$, there exists a function $\exp{J(x)}$. There are two local sections in a neighbourhood around $x$, identified with functions $\exp{J(x)}_i$ and $\exp{J(x)}_j$. It follows by construction that these are related by a left action of some $\Exp$ element, such that transition functions on $X$ are valued in $\Exp$. Thus, there is a fibre isomorphic to $\Exp$ over any point in $X$, defining a fibre bundle $P \xrightarrow{\pi_P} X$. Defining a commuting right action of $\Exp$ on the total space $P$, we take $\Exp$ as the structure group of a principal $\Exp$-bundle over $X$. Now, let $p : X \to \R$ and identify $s$ with $p$ as above. Suppose that there is a real vector bundle with a section that transforms $\Exp$-equivariantly under $\exp{J(x)} p(x)$. This is the identity representation of $\Exp$ such that $E \xrightarrow{\pi_E} X$ is associated to $\Exp$. The action given is that of a transition function for $p$ under the change in section $e^{J(x)}p(x)$, defining a probability density with a new set of constraints. By the fibre bundle construction theorem, $E$ is a vector bundle associated to $P$. 
\end{proof}

Note the obviously gauge-theoretic structure above: an arbitrary choice of probability density is possible (invariance), with a change in the density given by a corresponding change in the constraint (covariance). Likewise, a space which carries a representation of a Lie group generally has symmetry at the level of the space of solutions, but not for any particular solution. It is for this reason that principal and associated bundles are the natural mathematical language for gauge theory.

As we suggested was necessary after proving Theorem \ref{par-trans-at-max-ent-thm}, it is possible to associate a connection to this bundle.

\begin{prop}\label{connection-prop}
The vector of partial derivatives of $J$ comprises the set of coefficients of a gauge field. 
\end{prop}
\begin{proof}
Choose a coordinate chart for some neighbourhood of $X$, and let $\sigma : x \mapsto (x, g(x))$ be a local section $U \to P$. The pullback connection $\sigma^*\omega$ is a one-form $\Gamma(TU) \to T_eG$, that is, a Lie algebra-valued one-form on tangent vectors on the base. The pullback of $\omega$ is $A + g^{-1}\dd{g}$, an arbitrary one-form $A$ added to the Maurer--Cartan form $g^{-1} \dd{g}$. Set $A=0$ and observe that $\exp{-J(x)} \dd{\exp{J(x)}} = \partial_i J(x) \dd{x^i}$. The coefficients of this particular one-form are the components of $\grad J$ in the standard basis of $T\R^n$. 
\end{proof} 

The Maurer--Cartan form for $J$ is the choice of connection we will employ throughout the paper. Thus, from here on we understand $\sigma^* \omega = \dd{J}$ whenever we write either. The exactness condition on the connection is important, as we will see in the following two corollaries of Proposition \ref{connection-prop}:

\begin{corollary}\label{flat-bundle-cor}
Both bundles are flat and admit a covariantly constant section. 
\end{corollary}
\begin{proof}
Following Proposition \ref{connection-prop}, we have $\dd{J}$ for our connection one-form. Since this form is exact and is $\R$-valued, $\dd{(\dd{J})} + \dd{J} \wedge \dd{J} = 0$, implying the connection is flat. On a line bundle, a flat connection $A$ written locally as $\dd + \alpha$ has a nowhere vanishing parallel section if and only if $\alpha$ is exact. Proposition \ref{connection-prop} gives the desired result.
\end{proof}

\begin{corollary}\label{contour-cor}
The contour lines of $J$ are integral curves of the vector field $\ker(\sigma^*\omega)$.
\end{corollary}
\begin{proof}
Since the connection one-form is flat and is the exterior derivative of $J$, the horizontal distribution given by its kernel is involutive and its integral curves are level sets $J(x) = c$. 
\end{proof}

In particular, a consequence of Corollaries \ref{flat-bundle-cor} and \ref{contour-cor} is that parallel transport has a unique solution. 

\begin{remark}
Corollary \ref{contour-cor} is essential, since it tells us that the horizontal lifts that build our density are lifts of the contour lines of the surface $J$ as if they were projected down to $X$. This horizontal lift is identically parallel transport with respect to the gauge field, and also has the interpretation of weighting a state's probability inversely to the constraint on sampling that state. Viewing a constrained weighting as a constrained lift, we can appeal to the fact that the exponential function we compose $J$ with is simply the strictly positive function that preserves inverses (see the homomorphism given in Definition \ref{exp-defn}), making it a canonical choice for such a weighting. It moreover has the interpretation of a parallel transport map giving the horizontal lifts of the referenced contour lines.
\end{remark} 

We have previously mentioned that the gauge field determines the configuration of the gauge potential, and the dynamics of particles relative to the gauge potential are given by parallel transport along this connection. It follows that lifts which obey the connection are given by parallel transport. We denote the pullback of a connection by a local section using the local form $A \in \Omega^1(U; \g)$ for brevity. Suppose a particle moves along space-time on a path $\varphi$, and the connection along this path is given by the pre-composition $\varphi^* A = A(\varphi)$. For an abelian group $G$, the vector space isomorphism $\g \simeq T_e G$, and a local connection one-form $A$, the movement of a particle $\epsilon$ with initial state $(x_0, \,g(x_0) = g_0)$ through a gauge potential is given by the parallel transport equation
\begin{equation*}
\tra(\epsilon) = \exp\left\{ - \int_{[0,1]} \varphi^* A \right\}.
\end{equation*}
This is a geodesic curve in the fibres, following $A$ exactly, whose end-point is a horizontal lift $g(x_f)$ of $x_f$, parallel to the path in the base. 

\begin{prop}\label{par-trans-prop}
Setting $\sigma^* \omega = g^{-1} \dd{g}$ yields $\exp{J(x)}$ for $\traemph : \tilde\varphi(x_0) \mapsto \tilde\varphi(x)$.
\end{prop}
\begin{proof}
Let $\varphi : I \to X$ be a parameterised curve in $X$. For the map $\tra(\epsilon)$ transporting an object along the lift of the path $\varphi$, we have
\[
\exp{-\int_{I} \varphi^*A }.
\]
Under our gauge, and by properties of the pullback, we have
\[
-\int_{\varphi(0)}^{\varphi(1)} g^{-1} \dd{g} = -\int_{x_0}^x e^{-\lambda J(x)} \partial_i \lambda J(x) e^{\lambda J(x)} \dd{x^i}.
\]
Since this form is exact, the transport of a point above $x_0$ to a point above $x$ in parallel fashion is
\[
\exp{-\lambda J(x)}\exp{-\lambda J(x_0)}.
\]
As such, points on the surface parameterised by $(x, p(x))$ taken as end-points of some horizontal lift are given by $\exp{-\lambda J(x) }$.
\end{proof}

The claim is that taking $J$ as a gauge symmetry and asking that the probability of states obeys $J$ is asking that a point of probability, $p(x)$, transports itself over $X$ according to the constraints. To say that the flows or trajectories given by a horizontal lift\textemdash formally, lines on a surface of interest\textemdash follow the connection can be visualised as these lines evolving along the density and direction of the level sets of $J$, which themselves indicate how preferable a state is. To maximise entropy is to obey the information contained in the constraints on states. Note that maximum entropy assumes the generating process for states is stationary. The flow being referred to is not a description of the change in time, but roughly the change in state space, of the probability density resulting from $J$. 

This recovers the integrability result hinted at in Theorem \ref{par-trans-at-max-ent-thm}\textemdash one can imagine transporting some initial point $p(x)$ parallel to the contour lines given by the constraints, such that finding $p(x)$ is reduced to solving a linear equation describing the particular spatial shape of the density. That is, we take the probability density as a surface of dimension $\dim(X)$ embedded in the total space of $E$ by the section $s : x \mapsto (x, p(x))$, which consists of integral curves of a tangent vector field to that surface. This vector field is given by the horizontal tangent subbundle, such that the probability density consists of horizontal lifts of contour lines of the pullback one-form. Whilst verbose, the geometry is simple; contrast this with solving a PDE for diffusion in a potential by actually calculating the double integral we have avoided, and the advantage is obvious. There is a very direct sense in which we offload the complexity into the form of the differential operator $\omega$, but as this is a linear operator, that complexity can be circumvented by the general solution to matrix differential equations. As mentioned, parallel transport is the solution to this system. This is a similar trick to reducing a PDE to a one-parameter linear differential equation (by, for instance, the method of characteristics), found in elementary PDE analysis. 

\begin{example}\label{example}
Here we provide a worked example, namely, that of quadratic constraints on a two-dimensional real base. Let $\exp{\lambda J} : X \to \R$ correspond to the section 
\[
(x, y) \mapsto (x, y, \exp{x^2 + y^2}).
\]
The pullback connection has as components the two-dimensional tangent vector field consisting of vectors $(2x, 2y)$. Under a change in constraint, we have demonstrated this transforms like $A_i \mapsto A_i + \partial_i J(x, y) \dd{x^i}$, which adds the derivative of each vector component to the original connection. One can easily verify that this returns the connection of a new section. By Proposition \ref{par-trans-prop}, parallel transportation along this connection leads to\textemdash when we include a normalisation constraint and solve for its Lagrange multiplier\textemdash a two-dimensional Gaussian probability density centred at zero, 
\[
Z^{-1} \exp{- x^2 - y^2}.
\]
Likewise, the kernel of $\sigma^*\omega = 2x\dd{x} + 2y\dd{y}$ is the distribution on $\R^2 \setminus \{0\}$ generated by $-2y \partial_x + 2x \partial_y $ and $ 2y \partial_x - 2x  \partial_y $, which has integral curves $x^2 + y^2 = c$. When the pullback to the base is lifted into $E$, these integral curves are horizontal circles along which transport is parallel. Once more, this ultimately formalises the statement that the likelihood of a state is weighted by the constraints in a very specific fashion that determines how probability is placed over $X$. Each lift comprising the density is a horizontal level set of equiprobable states, which assemble into a surface by homotopy.
\end{example}

With that example being stated, we provide a theorem tying together all of our results so far.\footnote{The author is grateful to James F Glazebrook for suggesting part of the phrasing of Theorem \ref{main-result}, whose statement has been much improved.}

\begin{theorem}\label{main-result}
Maximum entropy inference arises from a geometry on the state space induced by a set of constraints, where those constraints are coupled to the measure over field states $p(x)$ in the field-theoretic sense of Theorem \ref{J-is-gauge-field-thm}. Consequently, a scalar field $p(x)$ is the stationary solution to some Fokker--Planck equation if and only if parallel transport solutions in that geometry exist.
\end{theorem}
\begin{proof}
We recall Theorem \ref{par-trans-at-max-ent-thm}, which established that, when $J$ is a connection, the stationary point of $S[p;J]$ is equivalent to parallel transport with respect to $J$. By Theorem \ref{J-is-gauge-field-thm}, $J$ is a gauge field for the action $S : \Gamma(E) \to \R$, with the interpretation of a pullback connection one-form on a principal $G$-bundle associated to $E$ by Proposition \ref{connection-prop}. By Corollary \ref{contour-cor}, this connection admits integral curves $J(x) = c$, whose horizontal lifts are given by the parallel transport map $\exp{-\lambda J(x)}$. Since the constraints specified are holonomic, the integral curve of any vector field in the distribution is the solution to a constrained equation of motion in the state space, whose lift gives the probability of observing any state along that curve.
Thus it is the case that maximising entropy builds a probability distribution by horizontally lifting contour lines of the pullback connection, such that the integral curves of the evolution of points $p(x)$ are horizontal paths on a surface implicitly described by $(x, p(x))$. By classic results in functional analysis, $p(x)$ is the equilibrium solution to a Fokker--Planck equation in a scalar potential $J(x)$ if and only if it satisfies the stationary point of a gradient flow on a free energy functional, which is identically a constrained entropy functional. Since, by Theorem \ref{par-trans-at-max-ent-thm}, the maximisation of entropy is parallel transport, $p(x)$ solves the Fokker--Planck equation in a potential if and only if parallel transport exists in the bundle described here.
\end{proof}

In effect, we have shown that maximising entropy lifts a sampling process in $X$ to a space of probabilities $E$, constrained by the connection, and that this is how probabilities are produced by the entropy functional. This is the main result of the paper.

\section{Consequences for statistical inference and physics}\label{consequences-sec}

We have made a somewhat bold claim in these results, which is that, in some sense, gauge theory is the `right' way of thinking about statistical inference and parts of functional analysis relevant to statistics in inference and physics. To summarise, we have viewed a probability density as a surface consisting of integral curves (and their respective $n$-dimensional generalisations) whose corresponding tangent vector fields are given by the distribution $\ker(\sigma^*\omega)$, or, the horizontal tangent space of $P$ pulled back to $X$. That this arises from a principal $G$-bundle with a connection form coupling to the covariant derivative on $E$ arises from the gauge-invariant nature of $S$ as a functional on field configurations $p$, and is reflected in the fact that the stationary point of $S$ with a fixed constraint $J$ is parallel transport with respect to $J$. This further reflects that the shape of this density is dictated by the geometry induced by the constraints, as in a matter field coupled to a gauge field. Indeed, the characterisation of the principle of maximum entropy as lifting integral curves of the constraint distribution reveals the very point of constrained maximum entropy: all we need to do to solve a problem is characterise it with a set of constraints, and then a simple horizontal lift gives us the sought probability density $p(x)$. Since we know the function $J$ \emph{a priori} and $\dd{J}$ is exact, these integral curves are easy to find; moreover, the lift is given by parallel transport, a linear problem whose solution exists uniquely and can be found non-perturbatively. 

We propose to call this effective geometry for dynamics in a space, induced by the constraints on those dynamics, a \emph{constraint geometry}. We believe this usefully encodes other properties of stochastic systems, which we will discuss here. 

Recall that, for an exact form $\sigma^*\omega = \dd{f}$ and a metric on $X$, the vector field dual to $\sigma^*\omega$ is the gradient of $f$\textemdash a field of vectors perpendicular to the vectors tangent to the contour lines of $f$. Indeed, these are lines for which the covector field $\sigma^*\omega$ is a linear approximation at each point. We can sketch an immediate consequence of the gauge-theoretic aspects of these results, for a class of inference algorithm that does estimation: crossing contour lines introduces a climbing component to the flow, which is given by non-zero vertical components of the vectors tangent to that flow. Moreover, this results in a splitting of the diffusion dynamics sampling from that density into two components, one vertical, and one horizontal.

\begin{prop}
Maximising likelihood is a charged flow. 
\end{prop}
\begin{proof}
By definition of parallel transport, horizontal flows are along lifts of paths $J(x) = c$, such that tangent vectors to these flows are horizontal. As a result, each horizontal lift is a non-empty locus of points $x$ of equal probability $\exp{-J(x)} = \exp{-c}$. Since maximising probability is equivalent to decreasing the value of level sets of $J(x)$, any such flow is not along a horizontal lift. A gradient climbing flow on $(x, p(x))$ is thus equivalent to a gauge force acting on that flow.
\end{proof}

In this sense, the geodesic curvatures of paths on the surface $(x, J(x))$, which encode the gauge force acting on paths on $(x, p(x))$, can be understood as facilitating the convergence of paths to areas of high probability. This suggests the curvature of the constraint surface as an important quantity with which to understand dynamics under a fixed choice of gauge. Indeed, when a vertical flow can be characterised by the change in the gradient dual to $\dd{J}$, we ought to be able to derive information about the dynamics of paths in $X$ and $E$. 



By splitting the tangent space $TP$ into horizontal and vertical components, we are able to induce a splitting of flows on $X$ into analogous components. Interestingly, decompositions of this type\textemdash having gradient flow and circulatory components\textemdash are known to exist for equilibrium and non-equilibrium steady state systems, the latter having invariant measures that break detailed balance \cite{graham, ao, hong, afepfapp, barp2}. More generally, the combination of dissipative and conservative flows characterises any invariant measure of Gibbs-type \cite{barp}, and work of It\=o \cite{ito} and Elworthy--Le Jan--Li \cite{elworthy2, kd-lang} reference similar horizontal-vertical decompositions. These results extend this characterisation of measure-preserving diffusion to a gauge-theoretic framework. 

In Remark \ref{differences-rem}, we discussed how this construction focusses on a stationary solution to the Fokker--Planck equation, as opposed to previous results that concern flows on a statistical manifold or non-equilibrium steady states with fluxes. Even given this key difference, there is some (clearly non-trivial) convergence of our results with previous work. In particular, \cite{polettini} derives a similar invariance for $p(x)$ as we have here, in an effort to understand constraints on non-equilibrium processes. Likewise, \cite{graham}\textemdash which is the genesis of the class of decomposition cited above\textemdash derives a similar result to ours which also resembles parallel transport. In fact, this is produced using a covariance argument under changes in coordinates\textemdash but not a gauge symmetry, explicitly. Hence, our results appear to generalise certain ideas in statistical physics in a way which is compatible with older results in the literature. A more complete account of this splitting will appear in forthcoming work. For now we only note that the stationary Fokker--Planck equation on a Riemannian manifold has generator
\[
- g^{ij}\Gamma_{ij}^k \partial_k + g^{ij}\partial_{ij} 
\]
where $g$ is the metric. Clearly this coincides with our results when the connection coefficients are given as in Example \ref{example}.

It is evident that solving the Fokker--Planck equation via the maximisation of entropy relies on parallel transport dynamics in `particle' space. Looking forward, a critical future direction would be to extend this work to dynamics in `distributional' space, to discover new geometric characterisations of dynamics in non-equilibrium regimes. Connections to existing frameworks for parallel transport on statistical manifolds \cite{friston} and Wasserstein spaces \cite{lott}, as well as time-dependent statistical mechanics \cite{presse}, could be accommodated by such a generalisation. 


\addtocontents{toc}{\SkipTocEntry}
\subsection*{Conflict of interest}

On behalf of all authors, the corresponding author states that there is no conflict of interest.

\addtocontents{toc}{\SkipTocEntry}
\subsection*{Data availability} No data generated.

\bibliographystyle{alpha}
\bibliography{main}

\end{document}